# Point Defects and Grain Boundaries in Rotationally Commensurate MoS$_2$ on Epitaxial Graphene

Xiaolong Liu[1], Itamar Balla[2], Hadallia Bergeron[2], and Mark C. Hersam[1,2,3,4*]

[1]Applied Physics Graduate Program, [2]Department of Materials Science and Engineering, [3]Department of Chemistry, [4]Department of Electrical Engineering and Computer Science, Northwestern University, Evanston, IL 60208, USA

*Correspondence should be addressed to: m-hersam@northwestern.edu, +1-847-491-2696

**ABSTRACT**

With reduced degrees of freedom, structural defects are expected to play a greater role in two-dimensional materials in comparison to their bulk counterparts. In particular, mechanical strength, electronic properties, and chemical reactivity are strongly affected by crystal imperfections in the atomically thin limit. Here, ultra-high vacuum (UHV) scanning tunneling microscopy (STM) and spectroscopy (STS) are employed to interrogate point and line defects in monolayer MoS$_2$ grown on epitaxial graphene (EG) at the atomic scale. Five types of point defects are observed with the majority species showing apparent structures that are consistent with vacancy and interstitial models. The total defect density is observed to be lower than MoS$_2$ grown on other substrates, and is likely attributed to the van der Waals epitaxy of MoS$_2$ on EG. Grain boundaries (GBs) with 30° and 60° tilt angles resulting from the rotational commensurability of MoS$_2$ on EG are more easily resolved by STM than atomic force microscopy at similar scales due to the enhanced contrast from their distinct electronic states. For example, band gap reduction to ~0.8 eV and ~0.5 eV is observed with STS for 30° and 60° GBs, respectively. In addition, atomic resolution STM images of these GBs are found to agree well with proposed structure models. This work offers quantitative insight into the structure and properties of common defects in MoS$_2$, and suggests pathways for tailoring the performance of MoS$_2$/graphene heterostructures *via* defect engineering.

**TOC Figure**

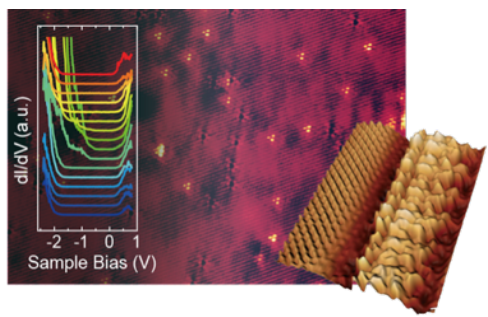

## 1. INTRODUCTION

The advent of graphene and the subsequent extensive research studying this superlative material[1-3] have demonstrated the unique opportunities and challenges associated with materials in the atomically thin limit. Similar to graphene, transition metal dichalcogenide (TMDC) monolayers[4] exhibit unique material properties compared to their bulk counterparts. As a result of the numerous combinations of transition metals with chalcogen elements,[5] the diverse family of TMDC compounds[6] offers electronic properties ranging from semiconducting (*e.g.*, $MoS_2$[7,8]), to metallic (*e.g.*, $NbTe_2$[9,10]) to superconducting (*e.g.*, $NbSe_2$[11,12]). In addition to being abundant in nature as molybdenite, $MoS_2$ is one of the most studied two-dimensional (2D) TMDCs with potential applications in electronics and optoelectronics due to its desirable physical properties.[13,14] Since monolayer $MoS_2$ possesses a direct optical band gap of 1.9 eV in the visible range[15] and a relatively high quantum yield,[8] it is especially promising for applications in photodetectors.[16,17] In particular, when combined with graphene, the resulting $MoS_2$/graphene heterostructure photodetectors show notably high gain.[17] The break of spin degeneracy due to the lack of inversion symmetry in monolayer $MoS_2$ also makes it possible to fabricate valleytronic devices controlled by circularly polarized light,[18,19] thereby offering further opportunities for quantum electronics.

Despite the aforementioned attractive properties of monolayer $MoS_2$, its relatively low experimentally observed carrier mobility[20] presents limitations in many high-performance applications. Previous work has suggested that intrinsic structural defects such as vacancies and grain boundaries (GBs) are the main sources of this degraded mobility.[21,22] Furthermore, the fact that $MoS_2$ field-effect transistors perform poorly in ambient compared to vacuum conditions implicates surface defects as sites of chemisorbed extrinsic scattering centers.[23,24] On the other hand, defect engineering presents pathways for modifying material properties, especially when the preparation of large-scale monolayer $MoS_2$ relies primarily on chemical vapor deposition (CVD)[25-28] and solution-phase exfoliation,[29,30] where high defect density is expected. For example, anisotropic defects in related 2D materials (e.g., graphene and black phosphorus[31]) are being explored in separation technologies as nanopores. Gate-tunable $MoS_2$ memristors based on GB migration[32] present another example where structural defects have been exploited to reveal novel functionalities.

A comprehensive understanding of $MoS_2$ structural defects at the atomic scale would facilitate the rational design of defect engineering for desired material properties. Although various studies of monolayer $MoS_2$ defects have been carried out by scanning/transmission electron microscopy (S/TEM),[26,33-38] concurrent structural and electronic characterization of monolayer $MoS_2$ defects with scanning tunneling microscopy and spectroscopy (STM/STS) is relatively unexplored.[39] Towards this end, we perform here ultra-high vacuum (UHV) STM/STS measurements of CVD-grown monolayer $MoS_2$ on epitaxial graphene (EG), focusing on the structural and electronic properties of intrinsic defects. Previously, we have demonstrated the growth of rotationally commensurate $MoS_2$ on EG,[25] which offers several technological advantages, including consistent material quality and preferential formation of 30° and 60° GBs. The investigation of point defects and GBs in the $MoS_2$/EG system is thus of particular interest because of the strain-free nature of the monolayer $MoS_2$ and the predictability in GB orientations enabled by the rotational commensurability.[40]

## 2. EXPERIMENTAL METHODS

**Synthesis of Epitaxial Graphene on SiC.** A 5 mm × 9 mm 4H-SiC (0001) (Cree Inc.) wafer was degassed at 550 °C under UHV conditions (~6 × 10$^{-10}$ Torr) for 12 hours and subsequently brought to 1270 °C at a rate of 100 °C/min. At this temperature, the graphitization process takes place by thermal desorption of Si leaving behind 1 to 2 layers of graphene. After 20 min at 1270 °C, the substrate was cooled down to room temperature at a rate of 50 °C/min.

**Synthesis of $MoS_2$ on Epitaxial Graphene.** The graphitized SiC wafer was placed 1 inch downstream (Si face up) of an alumina boat containing 10 mg of molybdenum trioxide powder ($MoO_3$, 99.98% trace metal, Sigma-Aldrich) located at the center of the heating zone of a 1 inch cylindrical quartz tube furnace (Lindberg/Blue). Another alumina boat containing 150 mg of sulfur powder (Sigma-Aldrich) was located 12 inches upstream of the $MoO_3$ boat. The temperature of the sulfur source was maintained using a heating belt wrapped around the quartz tube and controlled by a proportional-integral-derivative controller (Omega). The tube was then sealed and purged with argon gas to reach a base pressure of ~50 mTorr. Following the purge, the furnace was heated to 150 °C for 20 min (in order to desorb any residual water inside the tube). The furnace was then heated to 800 °C in 55 min, maintained at 800 °C for 15 min, and cooled down to room temperature naturally. The sulfur was initially maintained at 50 °C for 49 min (until the furnace reaches ~500 °C), brought to 120 °C in 17 min, maintained at 120 °C for 18 min, then cooled down to room temperature naturally. During all stages of heating and cooling, the tube was kept at 40 Torr under inert conditions using argon as a carrier gas at a flow rate of 25 sccm.

**Scanning Tunneling Microscopy and Spectroscopy.** All STM and STS measurements were carried out in a home-built UHV STM system[41] with the microscope based on the Lyding design.[42] The base pressures of the STM chamber and preparation chamber were $7 \times 10^{-11}$ Torr and $6 \times 10^{-10}$ Torr, respectively. After the $MoS_2$/EG sample was loaded through the loadlock chamber, it was degassed at ~205 °C for 6 hours prior to scanning. The electrochemically etched PtIr tip (Keysight) was grounded, and a bias voltage was applied to the sample. Control electronics and software from Nanonis were used for tip manipulation and data collection. The tip approach was monitored with a long working distance optical microscope (GT Vision). To acquire STS spectra, a lock-in amplifier (SRS model SR850) was used with an RMS amplitude of 30 mV and a modulation frequency of ~8.5 kHz.

**Atomic Force Microscopy.** Tapping mode atomic force microscopy (AFM) was performed on an Asylum Cypher AFM with NCHR-W Si cantilevers (NanoWorld). The resonant frequency of cantilevers was ~320 kHz, and the scanning rate was 1 Hz.

**Raman Spectroscopy.** Raman spectra were taken in ambient conditions with a Horiba Scientific XploRA PLUS Raman microscope. The 532 nm laser line was focused on the sample surface with a typical spot size of 1 μm and power of 1 mW. The scattered light was dispersed by a grating of 2400 grooves/mm. The spectra were acquired over 30 sec and averaged over 3 spectra to obtain an improved signal to noise ratio. Raman peak position calibration was performed against the SiC band[43] at 776.65 cm$^{-1}$.

## 3. RESULTS AND DISCUSSION

The as-grown $MoS_2$/EG sample was imaged by AFM as shown in Figure 1a. The vast majority of monolayer $MoS_2$ triangle domains are aligned as a result of rotational commensurability. The high quality of $MoS_2$ is revealed by Raman spectroscopy, as shown in Figure 1b, where the in-plane $E_{2g}^1$ (386.5 cm$^{-1}$) and out-of-plane $A_{1g}$ (407.1 cm$^{-1}$) Raman modes[44] are separated by 20.6 cm$^{-1}$, corresponding to monolayer $MoS_2$.[45] The STS-measured band gap of ~2 eV in Figure 1c further indicates good material quality.[25] Due to the large mismatch between the lattice constants of $MoS_2$ (3.16 Å) and EG (2.46 Å), which prevents direct epitaxy but allows van der Waals epitaxy of $MoS_2$, a Moiré superstructure is expected to develop.[46] As shown in Figure 1d, Moiré patterns with different levels of contrast but same periodicities are observed at different scanning conditions by UHV STM. To observe Moiré patterns, the tunneling current must be modulated by both the top $MoS_2$ and underlying EG layers concurrently, which implies that the appearance of Moiré patterns is dependent on tunneling conditions.[47] At certain tunneling conditions, Moiré patterns can disappear, as observed both in previous literature[48] and in our experiments (Figure S1). The relatively high tunneling current used in Figure 1d ensures an effective overlap between the electron wave functions from both the $MoS_2$ and EG layers, thus allowing STM observation of the Moiré patterns. A schematic illustration of the observed Moiré patterns is provided in Figure S2, which further confirms that the $MoS_2$ lattice is well-aligned with that of the EG.

Although the monolayer $MoS_2$ crystal domains are flat over relatively large length scales as imaged by AFM (Figure 1a), several types of point defects are present when examined at the atomic scale in STM. For example, Figure 2a is an STM topography image of the monolayer $MoS_2$ surface recorded at $V_{sample}$ = -1 V with the crystal lattice resolved and decorated with apparently bright and dark defects. Additional STM images of point defects are provided in Figure S3. Four types of atomic defects are identified and labeled as type-1 to type-4 with each indicated by a green, yellow, red, and white arrow, respectively. From Figure 2a, the majority of point defects can be characterized as type-1 and type-2, with respective areal densities of 4.2 × 10$^{12}$/cm$^2$ and 3.8 × 10$^{12}$/cm$^2$. Averaging over several images, the total defect density is ~8 × 10$^{12}$/cm$^2$, which is lower than the reported defect densities of $MoS_2$ directly grown or mechanically exfoliated onto $SiO_2$.[22,35] This low defect density can likely be attributed to the registry of MoS$_2$ with EG, and is consistent with the high crystal quality that is expected for van der Waals epitaxy.[49,50]

High-resolution STM images showing type-1 to type-4 defects are shown in Figure 2b and c with an additional type-5 defect shown in Figure 2d. The type-1 and type-2 defects have well-defined structures in which the type-1 defect appears as a depression and the type-2 defect consists of 3 bright protrusions. Chalcogen vacancy defects in TMDCs (e.g., sulfur vacancies in MoS$_2$[35] and selenium vacancies in TiSe$_2$[51]) are common, and their prevalence is explained by the low vacancy formation energy of chalcogen atoms as calculated by density functional theory (DFT), whereas metal vacancies are energetically unfavorable.[35,52,53] Hence, it is likely that type-1 defects are sulfur monovacancies, which is consistent with previous literature that assigned apparent depressions in STM images to chalcogen vacancies.[51,54,55]

Type-2 defects appear as aligned triangles formed by three point protrusions. Such structures have been experimentally observed in TiSe$_2$ and assigned to Ti interstitial atoms.[55] However, given the much higher formation energy of interstitial Mo defects in comparison to S interstitials,[52] type-2 defects are more likely to be S interstitials, which implies that the type-1 and type 2 defects are possibly Frenkel pairs. Although Frenkel defects in non-layered bulk materials are associated with high formation energy, the empty sites in the van der Waals gap between MoS$_2$ and EG may effectively lower the activation energy in this case.[56] Alternatively, considering the low formation energy of sulfur divacancies,[35] such defects can also be prevalent. Indeed, recent work[57] suggests that sulfur divacancies also appear as triangular protrusions in simulated STM images. Therefore, type-2 defects are also consistent with a sulfur divacancy model.

The type-3 and type-4 defects shown in Figure 2c do not have well-defined boundaries and vary in size (Figure 2a). The continuous MoS$_2$ lattice through the defect area indicates that the origin of these defects is likely to be subsurface (*e.g.*, defects in the underlying EG or the bottom sulfur layer in monolayer MoS$_2$). Finally, the type-5 defect in Figure 2d appears as an adsorbed particle, possibly originating from the CVD source materials. The depression surrounding the particle may result from an electron depletion zone, which can be caused by Coulomb repulsion from a negatively charged defect. In Figures 2e-g, the bias-dependent behavior of type-1 and type-2 defects are provided. At different biases, type-1 defects do not show obvious changes in appearance, whereas type-2 defects develop depressions at their centers.

In both filled and empty states images, type-1 defects appear as depressions, which suggests the presence of physical pits in topography as would be expected for sulfur vacancies.

A second class of structural defects in 2D materials are line defects, including dislocations and GBs. Compared to zero-dimensional point defects, which are primarily imaged with atomic resolution methods, one-dimensional GBs can be detected at larger length scales. Driven by the importance of understanding GBs in the 2D limit, a number of approaches have been developed to visualize $MoS_2$ GBs including preferential oxidation of GBs,[58] nonlinear optics,[59,60] and stacking of $MoS_2$ bilayers.[61] As a consequence of rotational commensurability, the GBs of $MoS_2$ grown on EG are restricted to tilt angles of 30° and 60°. Figure 3a is an AFM topography image of a 30° GB for monolayer $MoS_2$ on EG. The formation of such GBs involves a 30° rotated crystal domain as shown in the schematic of Figure 3b, where the angle between two triangular domain edges is 90°. Although domains with 30° rotated angles are less energetically favorable than aligned domains, the small fluctuations of binding energies with respect to rotation angle in van der Waals heterostructures implies that a secondary crystal orientation may still be present experimentally.[62] Because the 30° rotated domains comprise only 14% of the total crystals as quantified in our previous work,[25] 30° GBs are the minority species. While the 30° GB is weakly resolved by AFM as shown in Figure 3a and its inset, a GB of the same type shows stronger contrast in the STM topography image in Figure 3b, likely due to electronic effects that will be explored further later. In addition, Figure 3c shows an AFM image of a 60° GB with the inset emphasizing the GB region. A schematic and STM image of a 60° GB are shown in Figure 3d. These 60° GBs are twin GBs resulting from two triangular domains with opposite orientations.[34] Due to the three-fold symmetry of the $MoS_2$ lattice, the resulting GB has a 60° tilt angle. Again, with likely electronic contributions, the 60° GBs show significant contrast in STM, appearing as bright protrusions. Additional large-scale STM images of GBs are provided in Figure S4.

Further understanding of $MoS_2$ GBs is gained through atomic-resolution UHV STM imaging. In particular, Figure 4a shows a high-resolution STM image of a 30° GB, where the arrows denote the zigzag directions of the top (green arrow) and bottom (blue arrow) domains separated by the GB. Since GBs are bright protrusions in the filled states images (Figure 3b,d), height median matching is performed on Figure 4a so that both the domain interior regions and the GB structure show observable contrast in the image. The raw image is provided in Figure S6 to enable comparison of the bias-dependent apparent height. Unlike the disordered GBs found in CVD graphene where the GBs are curved,[63] most GBs in $MoS_2$ are observed to be straight. The orientation of the 30° GB in Figure 4a is denoted by the black arrow, which has a 45° rotation angle with respect to the zigzag direction of the top grain (green arrow). The GB possesses discontinuities and exhibits segments of ordered structure. The apparent width of the GB is ~2.3 nm, which is larger than the expected few lattice constant width from S/TEM.[26,33] However, the fact that constant current STM images are a convolution of real space and electronic structure makes STM imaging sensitive to local electronic properties in addition to atomic structure. The electronic disturbance resulting from the GB is expected to have a finite decay distance as discussed in Figure S5, which would lead to the GB appearing wider in STM imaging. The higher chemical reactivity of GBs may also result in adsorbed impurities that could contribute to the apparent width of the GBs.

A zoomed-in image of the region marked by the black rectangle in Figure 4a is shown in Figure 4b. A linear superstructure is found at the interface between the GB region and the bottom grain as indicated by the white dashed line. Along the GB direction, the periodicity of the superstructure is 4-fold great than that of the bottom lattice. Based upon this superstructure and the orientation of the GB, an atomic model is proposed in the bottom panel of Figure 4b. The green and blue arrows correspond to the zigzag lattice orientations of the top and bottom grains, respectively. The pink area represents the GB region bordered by periodic facets along the zigzag direction of the $MoS_2$ lattice. The periodicity of this superstructure is 4 times greater than that of the lattice, which is consistent with the STM image in the top panel of Figure 4b. The resulting GB orientation as indicated by the black arrow in the schematic is indeed rotated 45° from the zigzag direction of the top grain (green arrow). Hence, this atomic model satisfies the critical observed characteristics of the 30° GB. Bias-dependent images of this GB are provided in Figure S6, which show further details that support this proposed structure.

The electronic consequences induced by the 30° GB are explored with STS point spectra taken across the GB, as shown in Figure 4c. The distance between two adjacent points in the inset is 5 Å. Far from the GB region in both domains, the band gaps are uniformly ~2 eV, which is consistent with Figure 1c and literature reports of $MoS_2$ bandgaps on graphite substrates (2.15 eV,[64] 1.9 eV[65]). As the tip moves onto the GB, a band gap reduction to ~0.8 eV (*i.e.*, a reduction of ~1.2 eV) is observed and accomplished primarily by the rise of valence band maximum (VBM). This observation agrees qualitatively with the observed band gap reduction at MoS$_2$ random tilt GBs (*i.e.*, ~0.85 eV at a 18° GB) and the trend that GBs with higher angles possess smaller bandgaps.[39] Simulations and S/TEM results of GB structures also suggest that lattice distortions, point defects, and atoms with different coordination from the bulk lattice can be present in GBs.[21,26,33,66] Furthermore, the resulting higher chemical reactivity due to local strain (*e.g.*, from sulfur vacancies[67]) and/or under-coordinated atoms may lead to oxidation and adsorption of impurities at GBs. All of these factors may contribute to additional electronic states in the band gap and/or close to the band edges, leading to the reduced band gap at GBs.

A more common 60° GB is examined at the atomic scale in Figure 5. The green and blue arrows in Figure 5a represent the respective zigzag directions of the top and bottom grains rotated by 60° from each other. In this case, the GB has a rotation angle of 19° from the bottom grain (blue arrow) indicated by the black arrow. The width of this GB is approximately 2 nm, which is slightly smaller than that of the 30° GB. The zoomed-in periodic superstructure in Figure 5b shows a periodicity 3 times that of the MoS$_2$ lattice along the GB direction. The proposed atomic model includes similar faceted interfaces with the corresponding periodicity in the superlattice. The resulting 19° GB direction relative to the zigzag direction of the bottom grain within the model again matches the orientation measured experimentally by STM. Since STM topography images are a convolution of physical and electronic structure, determining the detailed atomic structure in the GB region is difficult. It has been proposed that a disordered transition region in monolayer MoS$_2$ exists to accommodate the local strain in a 18° tilt GB.[39] In addition, an STEM-based study[33] showed that 60° GBs (*i.e.*, twin GBs) are ~20° rotated from the zigzag direction of the MoS$_2$ lattice, which is consistent with the 19° rotation observed here. Our structure model is further confirmed by their GB structure model consisting of primarily 8-4-4 membered rings at the interface of the GB, and the two grains being faceted with the same structure proposed in Figure 5b.

Electronically, the 60° GB shows a slightly larger band gap reduction to ~0.5 eV primarily due to the rise of the VBM, as demonstrated by the series of STS spectra taken across the GB shown in Figure 5c. This trend is consistent with literature precedent where larger tilt angles result in smaller band gaps in monolayer MoS$_2$ on graphite.[39] The larger apparent height of the GB compared to grain interiors imaged at negative sample biases in Figure 5 can be explained by the higher differential tunneling conductance, which is proportional to the sample density of states (DOS), from -0.5 V to -2 V, when the tip is above the GBs. The increased DOS drives the tip farther away from the GBs to maintain a constant tunneling current during STM imaging. This electronic effect also explains why the apparent height of GBs decreases at positive biases as shown in Figure S6.

## 4. SUMMARY AND CONCLUSIONS

In summary, a detailed study of the intrinsic structural defects of rotationally commensurate CVD-grown monolayer $MoS_2$ on EG has been carried out at the atomic scale using STM and STS. Five types of point defects, including atomically resolved vacancies and interstitials, were observed and occurred at lower areal density than alternative monolayer $MoS_2$ sample preparation methods likely due to the nature of van der Waals epitaxy of $MoS_2$ on EG. GBs were found to be more clearly imaged by STM than AFM due to their pronounced electronic contrast. STS shows band gap narrowing at 30° and 60° GBs, resulting from additional DOS close to the VBM. Based on the relative orientations of the GBs and the periodicities of the observed superlattices, consistent structural models were proposed for each type of GB. Overall, this study provides fundamental insights and a basis for ongoing efforts to realize defect engineering in 2D TMDCs,[68,69] especially for commensurate heterostructures enabled by substrates like graphene where predictable GB orientations are present.

## ASSOCIATED CONTENT

**Supporting Information**

Supporting Information is available free of charge on the ACS Publications website, and includes atomic resolution images without Moiré patterns, schematic illustration of the formation of Moiré patterns, additional STM images of $MoS_2$ point defects, large-scale STM images of $MoS_2$ GBs, band profiles across 30° and 60° $MoS_2$ GBs, and bias-dependent images of $MoS_2$ GBs.

## AUTHOR INFORMATION

**Corresponding Author**

*E-mail: m-hersam@northwestern.edu

**Notes**

The authors declare no competing financial interest.

**ACKNOWLEDGMENTS**

The growth of MoS$_2$ by CVD was supported by the National Institute of Standards and Technology (NIST CHiMaD 70NANB14H012). UHV STM/STS measurements were supported by the U.S. Department of Energy SISGR program (DOE DE-FG02-09ER16109), Raman characterization was supported by the Office of Naval Research (ONR N00014-14-1-0669) with the Raman microscope funded by the Argonne-Northwestern Solar Energy Research (ANSER) Energy Frontier Research Center (DOE DE-SC0001059). The authors kindly thank Dr. Kan-Sheng Chen, Dr. Joshua Wood, Dr. Jian Zhu, and Andrew Mannix for valuable discussions.

**FIGURES**

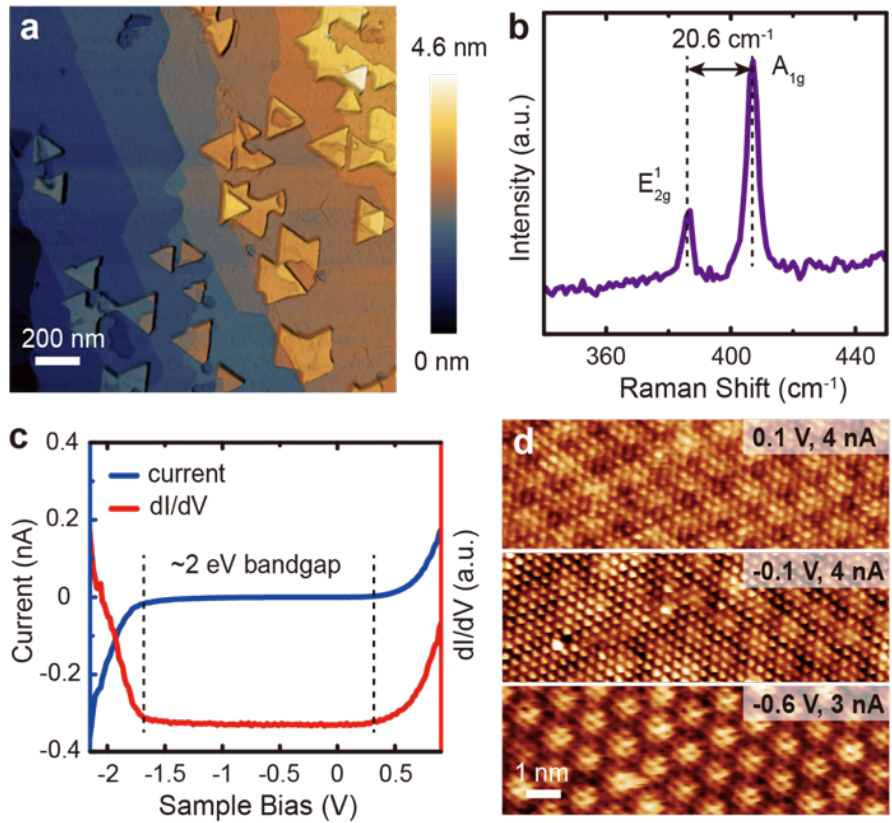

**Figure 1.** High crystal quality CVD-grown MoS$_2$ on EG. (a) AFM height image of predominantly monolayer MoS$_2$ domains on EG with aligned crystal orientations. (b) Raman spectrum of MoS$_2$ with E$_{2g}^1$ and A$_{1g}$ modes separated by 20.6 cm$^{-1}$. (c) Current and differential tunneling conductance spectra of monolayer MoS$_2$ on EG showing a band gap of 2 eV. (d) Moiré patterns from the MoS$_2$/EG heterostructure probed at different scanning conditions.

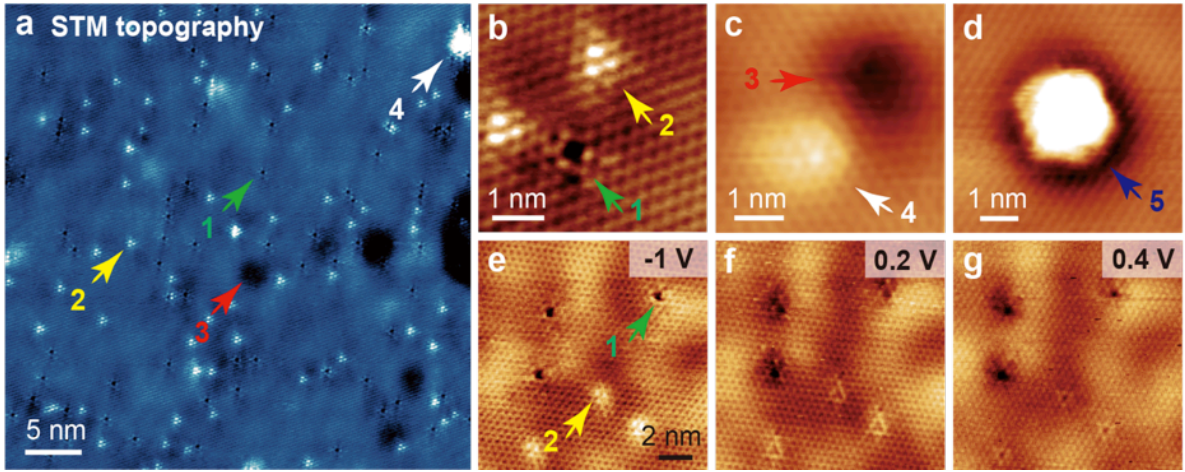

**Figure 2.** Point defects in monolayer MoS$_2$. (a) An STM topography image showing 4 types of point defects as indicated by the green, yellow, red, and white arrows. V$_{sample}$ = -1 V, I$_{tunneling}$ = 800 pA. (b-d) Atomic-scale imaging of representative point defects with an additional type-5 defect indicated by the blue arrow in (d). V$_{sample}$ = -1 V, I$_{tunneling}$ = 800 pA. (e-g) Bias-dependent images of type-1 and type-2 defects. I$_{tunneling}$ = 800 pA.

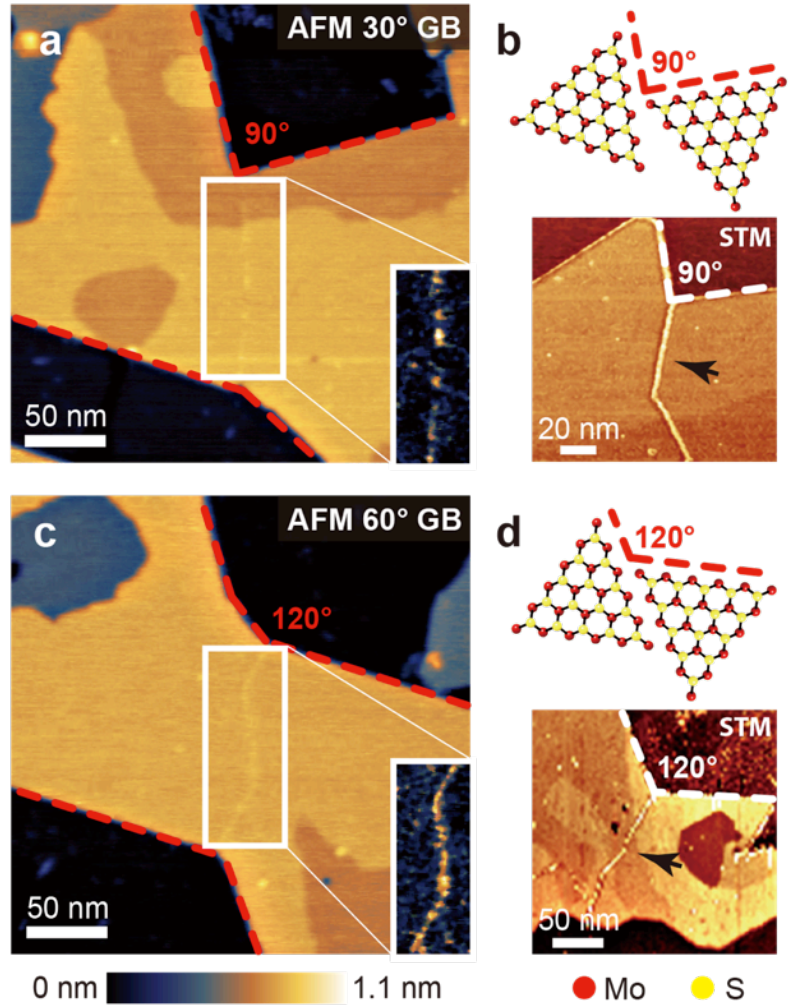

**Figure 3.** 30° and 60° GBs for rotationally commensurate monolayer MoS$_2$ on EG. (a,c) AFM height images of 30° and 60° GBs with the GB regions emphasized in the insets, respectively. The angles between the intersecting edges for the two types of GBs are 90° and 120°, respectively. (b,d) Schematics and STM topography images of 30° and 60° GBs. The GBs are indicated by black arrows in the STM images. V$_{sample}$ = -1 V, I$_{tunneling}$ = 50 pA.

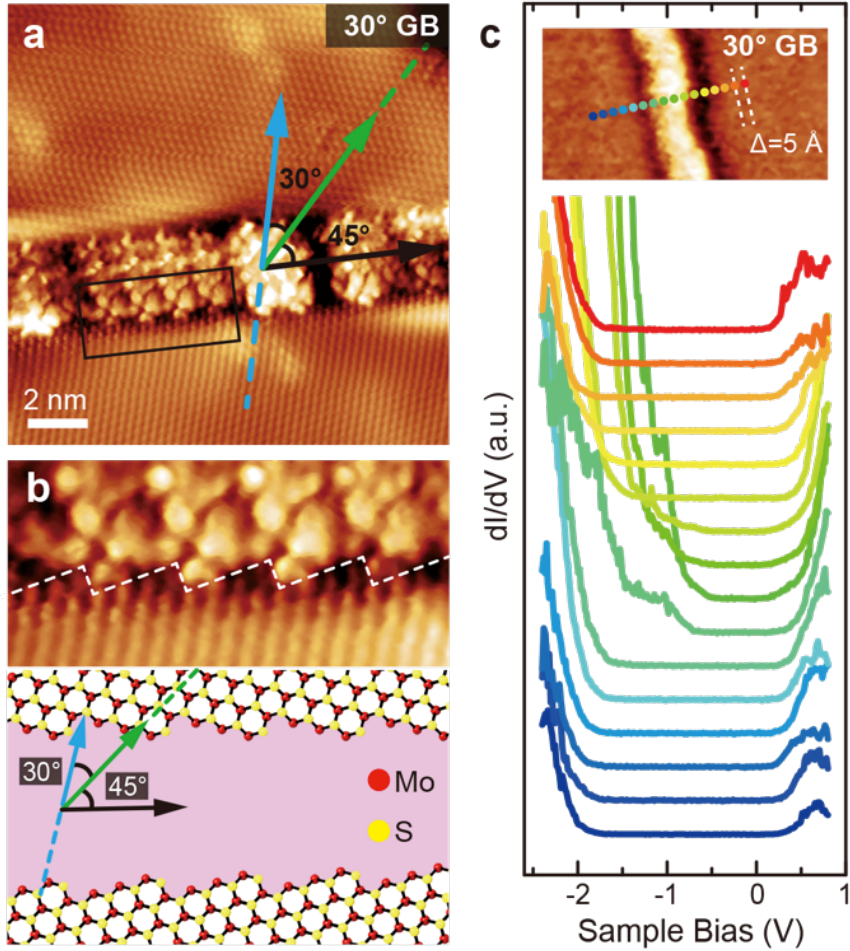

**Figure 4.** Atomic and electronic properties of 30° GBs. (a) An atomically resolved STM topography image of a 30° GB with the contrast adjusted by height median matching. The zigzag directions of the top and bottom grains are indicated by the green and blue arrows, respectively. The GB direction is indicated by the black arrow, which is rotated 45° from the zigzag direction of the top grain. $V_{sample}$ = -0.6 V, $I_{tunneling}$ = 5.2 nA. (b) Top: a zoomed-in STM image of the GB region marked by the rectangle in (a). Bottom: a structural model for the 30° GB with the green, blue, and black arrows corresponding to the zigzag directions of the top and bottom grains and the GB direction, respectively. (c). A series of STS spectra taken across a 30° GB with tip positions marked by the colored dots in the inset. The STS spectra show band gap narrowing across the GB. Adjacent points are separated by 5 Å.

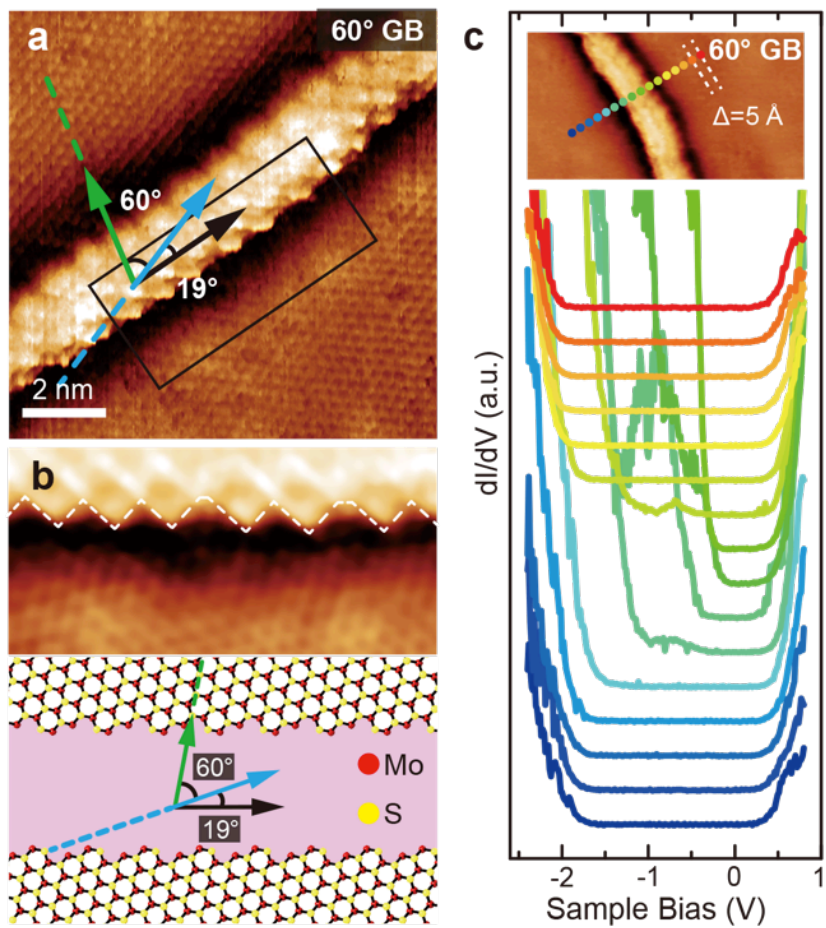

**Figure 5.** Atomic and electronic properties of 60° GBs. (a) An atomically resolved STM topography image of a 60° GB. The zigzag directions of the top and bottom grains are indicated by the green and blue arrows, respectively. The GB direction is indicated by the black arrow, which is 19° rotated from the zigzag direction of the bottom grain. $V_{sample}$ = -0.8 V, $I_{tunneling}$ = 50 pA. (b) Top: a zoomed-in STM image of the GB region marked by the rectangle in (a). A fast Fourier transform was applied to this image to better emphasize the superstructure of the GB edge. Bottom: a structural model for the 60° GB with the green, blue, and black arrows corresponding to the zigzag directions of the top and bottom grains and the GB direction, respectively. (c). A series of STS spectra taken across a 60° GB with tip positions marked by the colored dots in the inset. The STS spectra show band gap narrowing across the GB. Adjacent points are separated by 5 Å.

# REFERENCES

(1) Geim, A. K.; Novoselov, K. S. The Rise of Graphene. *Nat. Mater.* **2007**, *6*, 183–191.
(2) Secor, E. B.; Prabhumirashi, P. L.; Puntambekar, K.; Geier, M. L.; Hersam, M. C. Inkjet Printing of High Conductivity, Flexible Graphene Patterns. *J. Phys. Chem. Lett.* **2013**, *4*, 1347–1351.
(3) Xu, C.; Wang, X.; Zhu, J. Graphene–Metal Particle Nanocomposites. *J. Phys. Chem. C* **2008**, *112*, 19841–19845.
(4) Jariwala, D.; Sangwan, V. K.; Lauhon, L. J.; Marks, T. J.; Hersam, M. C. Emerging Device Applications for Semiconducting Two-Dimensional Transition Metal Dichalcogenides. *ACS Nano* **2014**, *8*, 1102–1120.
(5) Chhowalla, M.; Shin, H. S.; Eda, G.; Li, L.-J.; Loh, K. P.; Zhang, H. The Chemistry of Two-Dimensional Layered Transition Metal Dichalcogenide Nanosheets. *Nat. Chem.* **2013**, *5*, 263–275.
(6) Geim, A. K.; Grigorieva, I. V. Van Der Waals Heterostructures. *Nature* **2013**, *499*, 419–425.
(7) Ganatra, R.; Zhang, Q. Few-Layer $MoS_2$: A Promising Layered Semiconductor. *ACS Nano* **2014**, *8*, 4074–4099.
(8) Splendiani, A.; Sun, L.; Zhang, Y.; Li, T.; Kim, J.; Chim, C.-Y.; Galli, G.; Wang, F. Emerging Photoluminescence in Monolayer $MoS_2$. *Nano Lett.* **2010**, *10*, 1271–1275.
(9) Cukjati, D.; Prodan, A.; Jug, N.; van Midden, H. J. P.; Starowicz, P.; Karič, E.; Hla, S.-W.; Böhm, H.; Boswell, F. W.; Bennett, J. C. The Surface and Domain Structure of $NbTe_2$. *J. Cryst. Growth* **2002**, *237-239*, 278–282.
(10) Nagata, S.; Abe, T.; Ebisu, S.; Ishihara, Y.; Tsutsumi, K. Superconductivity in the Metallic Layered Compound $NbTe_2$. *J. Phys. Chem. Solids* **1993**, *54*, 895–899.
(11) Yokoya, T.; Kiss, T.; Chainani, A.; Shin, S.; Nohara, M.; Takagi, H. Fermi Surface Sheet-Dependent Superconductivity in $2H-NbSe_2$. *Science* **2001**, *294*, 2518–2520.
(12) Xi, X.; Wang, Z.; Zhao, W.; Park, J.-H.; Law, K. T.; Berger, H.; Forró, L.; Shan, J.; Mak, K. F. Ising Pairing in Superconducting $NbSe_2$ Atomic Layers. *Nat. Phys.* **2016**, *12*, 139-143.
(13) Ji, Q.; Zhang, Y.; Gao, T.; Zhang, Y.; Ma, D.; Liu, M.; Chen, Y.; Qiao, X.; Tan, P.-H.; Kan, M.; *et al*. Epitaxial Monolayer $MoS_2$ on Mica with Novel Photoluminescence. *Nano Lett.* **2013**, *13*, 3870–3877.
(14) Jariwala, D.; Sangwan, V. K.; Wu, C.-C.; Prabhumirashi, P. L.; Geier, M. L.; Marks, T. J.; Lauhon, L. J.; Hersam, M. C. Gate-Tunable Carbon Nanotube-$MoS_2$ Heterojunction p-n Diode. *Proc. Natl. Acad. Sci. U.S.A.* **2013**, *110*, 18076–18080.
(15) Mak, K. F.; Lee, C.; Hone, J.; Shan, J.; Heinz, T. F. Atomically Thin $MoS_2$: A New Direct-Gap Semiconductor. *Phys. Rev. Lett.* **2010**, *105*, 136805.
(16) Lopez-Sanchez, O.; Lembke, D.; Kayci, M.; Radenovic, A.; Kis, A. Ultrasensitive Photodetectors Based on Monolayer $MoS_2$. *Nat. Nanotechnol.* **2013**, *8*, 497–501.
(17) Zhang, W.; Chuu, C.-P.; Huang, J.-K.; Chen, C.-H.; Tsai, M.-L.; Chang, Y.-H.; Liang, C.-T.; Chen, Y.-Z.; Chueh, Y.-L.; He, J.-H.; *et al*. Ultrahigh-Gain Photodetectors Based on Atomically Thin Graphene-$MoS_2$ Heterostructures. *Sci. Rep.* **2014**, *4*, 3826.
(18) Mak, K. F.; McGill, K. L.; Park, J.; McEuen, P. L. The Valley Hall Effect in $MoS_2$ Transistors. *Science* **2014**, *344*, 1489–1492.
(19) Zeng, H.; Dai, J.; Yao, W.; Di Xiao; Cui, X. Valley Polarization in $MoS_2$ Monolayers by Optical Pumping. *Nat. Nanotechnol.* **2012**, *7*, 490–493.

(20) Novoselov, K. S.; Jiang, D.; Schedin, F.; Booth, T. J.; Khotkevich, V. V.; Morozov, S. V.; Geim, A. K. Two-Dimensional Atomic Crystals. *Proc. Natl. Acad. Sci. U.S.A.* **2005**, *102*, 10451–10453.

(21) Enyashin, A. N.; Bar-Sadan, M.; Houben, L.; Seifert, G. Line Defects in Molybdenum Disulfide Layers. *J. Phys. Chem. C* **2013**, *117*, 10842–10848.

(22) Qiu, H.; Xu, T.; Wang, Z.; Ren, W.; Nan, H.; Ni, Z.; Chen, Q.; Yuan, S.; Miao, F.; Song, F.; *et al*. Hopping Transport Through Defect-Induced Localized States in Molybdenum Disulphide. *Nat. Commun.* **2013**, *4*, 2442.

(23) Qiu, H.; Pan, L.; Yao, Z.; Li, J.; Shi, Y.; Wang, X. Electrical Characterization of Back-Gated Bi-layer $MoS_2$ Field-Effect Transistors and the Effect of Ambient on Their Performances. *Appl. Phys. Lett.* **2012**, *100*, 123104.

(24) Park, W.; Park, J.; Jang, J.; Lee, H.; Jeong, H.; Cho, K.; Hong, S.; Lee, T. Oxygen Environmental and Passivation Effects on Molybdenum Disulfide Field Effect Transistors. *Nanotechnology* **2013**, *24*, 095202.

(25) Liu, X.; Balla, I.; Bergeron, H.; Campbell, G. P.; Bedzyk, M. J.; Hersam, M. C. Rotationally Commensurate Growth of $MoS_2$ on Epitaxial Graphene. *ACS Nano* **2016,** *10*, 1067-1075.

(26) Najmaei, S.; Liu, Z.; Zhou, W.; Zou, X.; Shi, G.; Lei, S.; Yakobson, B. I.; Idrobo, J.-C.; Ajayan, P. M.; Lou, J. Vapour Phase Growth and Grain Boundary Structure of Molybdenum Disulphide Atomic Layers. *Nat. Mater.* **2013**, *12*, 754–759.

(27) Shi, Y.; Zhou, W.; Lu, A.-Y.; Fang, W.; Lee, Y.-H.; Hsu, A. L.; Kim, S. M.; Kim, K. K.; Yang, H. Y.; Li, L.-J.; *et al*. Van der Waals Epitaxy of $MoS_2$ Layers Using Graphene as Growth Templates. *Nano Lett*. **2012**, *12*, 2784–2791.

(28) Lin, Y.-C.; Lu, N.; Perea-Lopez, N.; Li, J.; Lin, Z.; Peng, X.; Lee, C. H.; Sun, C.; Calderin, L.; Browning, P. N.; *et al*. Direct Synthesis of van der Waals Solids. *ACS Nano* **2014**, *8*, 3715–3723.

(29) Kang, J.; Seo, J.-W. T.; Alducin, D.; Ponce, A.; Yacaman, M. J.; Hersam, M. C. Thickness Sorting of Two-Dimensional Transition Metal Dichalcogenides *via* Copolymer-Assisted Density Gradient Ultracentrifugation. *Nat. Commun.* **2014**, *5*, 5478.

(30) Varrla, E.; Backes, C.; Paton, K. R.; Harvey, A.; Gholamvand, Z.; McCauley, J.; Coleman, J. N. Large-Scale Production of Size-Controlled $MoS_2$ Nanosheets by Shear Exfoliation. *Chem. Mater*. **2015**, *27*, 1129–1139.

(31) Liu, X.; Wood, J. D.; Chen, K.-S.; Cho, E.; Hersam, M. C. *In Situ* Thermal Decomposition of Exfoliated Two-Dimensional Black Phosphorus. *J. Phys. Chem. Lett*. **2015**, *6*, 773–778.

(32) Sangwan, V. K.; Jariwala, D.; Kim, I. S.; Chen, K.-S.; Marks, T. J.; Lauhon, L. J.; Hersam, M. C. Gate-Tunable Memristive Phenomena Mediated by Grain Boundaries in Single-Layer $MoS_2$. *Nat. Nanotechnol*. **2015**, *10*, 403–406.

(33) van der Zande, A. M.; Huang, P. Y.; Chenet, D. A.; Berkelbach, T. C.; You, Y.; Lee, G.-H.; Heinz, T. F.; Reichman, D. R.; Muller, D. A.; Hone, J. C. Grains and Grain Boundaries in Highly Crystalline Monolayer Molybdenum Disulphide. *Nat. Mater*. **2013**, *12*, 554–561.

(34) Lin, J.; Pantelides, S. T.; Zhou, W. Vacancy-Induced Formation and Growth of Inversion Domains in Transition-Metal Dichalcogenide Monolayer. *ACS Nano* **2015**, *9*, 5189–5197.

(35) Hong, J.; Hu, Z.; Probert, M.; Li, K.; Lv, D.; Yang, X.; Gu, L.; Mao, N.; Feng, Q.; Xie,

L.; *et al*. Exploring Atomic Defects in Molybdenum Disulphide Monolayers. *Nat. Commun.* **2015**, *6*, 6293.
(36) Lin, Y.-C.; Björkman, T.; Komsa, H.-P.; Teng, P.-Y.; Yeh, C.-H.; Huang, F.-S.; Lin, K.-H.; Jadczak, J.; Huang, Y.-S.; Chiu, P.-W.; *et al*. Three-Fold Rotational Defects in Two-Dimensional Transition Metal Dichalcogenides. *Nat. Commun.* **2015**, *6*, 6736.
(37) Liu, Z.; Amani, M.; Najmaei, S.; Xu, Q.; Zou, X.; Zhou, W.; Yu, T.; Qiu, C.; Birdwell, A. G.; Crowne, F. J.; *et al*. Strain and Structure Heterogeneity in $MoS_2$ Atomic Layers Grown by Chemical Vapour Deposition. *Nat. Commun.* **2014**, *5*, 5246.
(38) Zhou, W.; Zou, X.; Najmaei, S.; Liu, Z.; Shi, Y.; Kong, J.; Lou, J.; Ajayan, P. M.; Yakobson, B. I.; Idrobo, J.-C. Intrinsic Structural Defects in Monolayer Molybdenum Disulfide. *Nano Lett.* **2013**, *13*, 2615–2622.
(39) Huang, Y. L.; Chen, Y.; Zhang, W.; Quek, S. Y.; Chen, C.-H.; Li, L.-J.; Hsu, W.-T.; Chang, W.-H.; Zheng, Y. J.; Chen, W.; *et al*. Bandgap Tunability at Single-Layer Molybdenum Disulphide Grain Boundaries. *Nat. Commun.* **2015**, *6*, 6298.
(40) Robinson, J. A. Growing Vertical in the Flatland. *ACS Nano* **2016**, *10*, 42-45.
(41) Foley, E. T.; Yoder, N. L.; Guisinger, N. P.; Hersam, M. C. Cryogenic Variable Temperature Ultrahigh Vacuum Scanning Tunneling Microscope for Single Molecule Studies on Silicon Surfaces. *Rev. Sci. Instrum.* **2004**, *75*, 5280–5287.
(42) Brockenbrough, R. T.; Lyding, J. W. Inertial Tip Translator for a Scanning Tunneling Microscope. *Rev. Sci. Instrum.* **1993**, *64*, 2225–2228.
(43) Lin, S.; Chen, Z.; Li, L.; Yang, C. Effect of Impurities on the Raman Scattering of 6H-SiC Crystals. *Mater. Res.* **2012**, *15*, 833–836.
(44) Lee, C.; Yan, H.; Brus, L. E.; Heinz, T. F.; Hone, J.; Ryu, S. Anomalous Lattice Vibrations of Single- and Few-Layer $MoS_2$. *ACS Nano* **2010**, *4*, 2695–2700.
(45) Plechinger, G.; Mann, J.; Preciado, E.; Barroso, D.; Nguyen, A.; Eroms, J.; Schüller, C.; Bartels, L.; Korn, T. A Direct Comparison of CVD-Grown and Exfoliated $MoS_2$ Using Optical Spectroscopy. *Semicond. Sci. Technol.* **2014**, *29*, 064008.
(46) Ugeda, M. M.; Bradley, A. J.; Shi, S.-F.; da Jornada, F. H.; Zhang, Y.; Qiu, D. Y.; Ruan, W.; Mo, S.-K.; Hussain, Z.; Shen, Z.-X.; *et al*. Giant Bandgap Renormalization and Excitonic Effects in a Monolayer Transition Metal Dichalcogenide Semiconductor. *Nat. Mater.* **2014**, *13*, 1091–1095.
(47) Hiebel, F.; Mallet, P.; Magaud, L.; Veuillen, J. Y. Atomic and Electronic Structure of Monolayer Graphene on 6H-SiC(000-1)(3×3): A Scanning Tunneling Microscopy Study. *Phys. Rev. B* **2009**, *80*, 235429.
(48) Feng, J.; Wagner, S. R.; Zhang, P. Interfacial Coupling and Electronic Structure of Two-Dimensional Silicon Grown on the Ag(111) Surface at High Temperature. *Sci. Rep.* **2015**, *5*, 10310.
(49) Koma, A.; Yoshimura, K. Ultrasharp Interfaces Grown with van der Waals Epitaxy. *Surf. Sci.* **1986**, *174*, 556–560.
(50) Koma, A. Van der Waals Epitaxy for Highly Lattice-Mismatched Systems. *J. Cryst. Growth* **1999**, *201–202*, 236–241.
(51) Peng, J.-P.; Guan, J.-Q.; Zhang, H.-M.; Song, C.-L.; Wang, L.; He, K.; Xue, Q.-K.; Ma, X.-C. Molecular Beam Epitaxy Growth and Scanning Tunneling Microscopy Study of $TiSe_2$ Ultrathin Films. *Phys. Rev. B* **2015**, *91*, 121113(R).
(52) Noh, J.-Y.; Kim, H.; Kim, Y.-S. Stability and Electronic Structures of Native Defects in Single-Layer $MoS_2$. *Phys. Rev. B* **2014**, *89*, 205417.

(53) KC, S.; Longo, R. C.; Addou, R.; Wallace, R. M.; Cho, K. Impact of Intrinsic Atomic Defects on the Electronic Structure of $MoS_2$ Monolayers. *Nanotechnology* **2014**, *25*, 375703.

(54) Addou, R.; Colombo, L.; Wallace, R. M. Surface Defects on Natural $MoS_2$. *ACS Appl. Mater. Interfaces* **2015**, *7*, 11921–11929.

(55) Hildebrand, B.; Didiot, C.; Novello, A. M.; Monney, G.; Scarfato, A.; Ubaldini, A.; Berger, H.; Bowler, D. R.; Renner, C.; Aebi, P. Doping Nature of Native Defects in 1T−$TiSe_2$. *Phys. Rev. Lett.* **2014**, *112*, 197001.

(56) Pehlke, E.; Schattke, W. The Effect of Frenkel Defects on the Electronic Structure of 1T-$TiSe_2$. *Z. Phys. B - Condensed Matter* **1987**, *66*, 31–37.

(57) González, C.; Biel, B.; Dappe, Y. J. Theoretical Characterisation of Point Defects on a $MoS_2$ Monolayer by Scanning Tunnelling Microscopy. *Nanotechnology* **2016**, *27*, 105702.

(58) Rong, Y.; He, K.; Pacios, M.; Robertson, A. W.; Bhaskaran, H.; Warner, J. H. Controlled Preferential Oxidation of Grain Boundaries in Monolayer Tungsten Disulfide for Direct Optical Imaging. *ACS Nano* **2015**, *9*, 3695–3703.

(59) Yin, X.; Ye, Z.; Chenet, D. A.; Ye, Y.; O'Brien, K.; Hone, J. C.; Zhang, X. Edge Nonlinear Optics on a $MoS_2$ Atomic Monolayer. *Science* **2014**, *344*, 488–490.

(60) Cheng, J.; Jiang, T.; Ji, Q.; Zhang, Y.; Li, Z.; Shan, Y.; Zhang, Y.; Gong, X.; Liu, W.; Wu, S. Kinetic Nature of Grain Boundary Formation in as-Grown $MoS_2$ Monolayers. *Adv. Mater.* **2015**, *27*, 4069–4074.

(61) Park, S.; Kim, M. S.; Kim, H.; Lee, J.; Han, G. H.; Jung, J.; Kim, J. Spectroscopic Visualization of Grain Boundaries of Monolayer Molybdenum Disulfide by Stacking Bilayers. *ACS Nano* **2015**, *9*, 11042–11048.

(62) Ji, Q.; Kan, M.; Zhang, Y.; Guo, Y.; Ma, D.; Shi, J.; Sun, Q.; Chen, Q.; Zhang, Y.; Liu, Z. Unravelling Orientation Distribution and Merging Behavior of Monolayer $MoS_2$ Domains on Sapphire. *Nano Lett.* **2015**, *15*, 198–205.

(63) Koepke, J. C.; Wood, J. D.; Estrada, D.; Ong, Z.-Y.; He, K. T.; Pop, E.; Lyding, J. W. Atomic-Scale Evidence for Potential Barriers and Strong Carrier Scattering at Graphene Grain Boundaries: A Scanning Tunneling Microscopy Study. *ACS Nano* **2013**, *7*, 75–86.

(64) Zhang, C.; Johnson, A.; Hsu, C.-L.; Li, L.-J.; Shih, C.-K. Direct Imaging of Band Profile in Single Layer $MoS_2$ on Graphite: Quasiparticle Energy Gap, Metallic Edge States, and Edge Band Bending. *Nano Lett.* **2014**, *14*, 2443–2447.

(65) Lu, C.-I.; Butler, C. J.; Huang, J.-K.; Hsing, C.-R.; Yang, H.-H.; Chu, Y.-H.; Luo, C.-H.; Sun, Y.-C.; Hsu, S.-H.; Yang, K.-H. O.; *et al.* Graphite Edge Controlled Registration of Monolayer $MoS_2$ Crystal Orientation. *Appl. Phys. Lett.* **2015**, *106*, 181904.

(66) Zou, X.; Liu, Y.; Yakobson, B. I. Predicting Dislocations and Grain Boundaries in Two-Dimensional Metal-Disulfides from the First Principles. *Nano Lett.* **2013**, *13*, 253–258.

(67) Li, H.; Tsai, C.; Koh, A. L.; Cai, L.; Contryman, A. W.; Fragapane, A. H.; Zhao, J.; Han, H. S.; Manoharan, H. C.; Abild-Pedersen, F.; *et al.* Activating and Optimizing $MoS_2$ Basal Planes for Hydrogen Evolution Through the Formation of Strained Sulphur Vacancies. *Nat. Mater.* **2016**, *15*, 48–53.

(68) Kim, I. S.; Sangwan, V. K.; Jariwala, D.; Wood, J. D.; Park, S.; Chen, K.-S.; Shi, F.; Ruiz-Zepeda, F.; Ponce, A.; Jose-Yacaman, M.; *et al.* Influence of Stoichiometry on the Optical and Electrical Properties of Chemical Vapor Deposition Derived $MoS_2$. *ACS Nano* **2014**, *8*, 10551–10558.

(69) Amani, M.; Lien, D. H.; Kiriya, D.; Xiao, J.; Azcatl, A.; Noh, J.; Madhvapathy, S. R.; Addou, R.; KC, S.; Dubey, M.; *et al*. Near-Unity Photoluminescence Quantum Yield in MoS$_2$. *Science* **2015**, *350*, 1065–1068.